\documentclass[reprint,showpacs,preprintnumbers, amsmath,amssymb,aps,prl,floatfix]{revtex4-1}

\usepackage{graphicx,color}

\usepackage{color,soul}
\usepackage{hhline}
\usepackage{dcolumn}
\usepackage{ amssymb }
\usepackage{bm}
\usepackage[left,modulo]{lineno} 
\usepackage{scrextend}
\usepackage[hidelinks]{hyperref}

\newcommand{\be}{\begin{equation}}
\newcommand{\ee}{\end{equation}}
\newcommand{\bea}{\begin{eqnarray}}
\newcommand{\eea}{\end{eqnarray}}

\begin{document}


\title{Neurofilaments function as shock absorbers: compression response arising from disordered proteins}

\author{Micha Kornreich, Eti Malka-Gibor, Ben Zuker, Adi Laser-Azogui, Roy Beck}
\email{roy@post.tau.ac.il}
\affiliation{Raymond and Beverly Sackler School of Physics and Astronomy\\ Tel Aviv
University, Ramat Aviv, Tel Aviv 69978, Israel}

\begin{abstract}

What can cells gain by using disordered, rather than folded, proteins in the architecture of their skeleton? Disordered proteins take multiple co-existing conformations, and often contain segments which act as random-walk-shaped polymers. Using X-ray scattering we measure the compression response of disordered protein hydrogels, which are the main stress-responsive component of neuron cells. We find that at high compression their mechanics are dominated by gas-like steric and ionic repulsions. At low compression, specific attractive interactions dominate. This is demonstrated by the considerable hydrogel expansion induced by the truncation of critical short protein segments. Accordingly, the floppy disordered proteins form a weakly cross-bridged hydrogel, and act as shock absorbers that sustain large deformations without failure. 


\end{abstract}

\maketitle

In the past two decades it was found that approximately 50\% of human proteins contain long disordered regions. The lack of specific structure has been shown to be critical for the disordered protein's designated functions \cite{Uversky2011,Almagor2013a}. As the conformation of these regions under physiological conditions resembles that of a random walk polymer, their investigation naturally calls for experimental and theoretical tools taken from polymer physics and statistical mechanics. 

Treating disordered proteins as polymers implies that their statistical conformations in space would not significantly change due to minor local modifications. Accordingly, if a modification of only a few specific amino acids (\textit{e.g.} monomers) does result in significant conformational changes, specific interactions between amino-acids must be involved \cite{Uversky2011,Kornreich2015a}. Identification of such sequence motifs in disordered proteins poses a challenge which is not met by current mean-field approaches of polymer physics. 

Here we experimentally identify motifs that dramatically alter the conformation of disordered proteins. These motifs locally cross-bridge protein segments, leading to the breakdown of the mean-field polymer approach. We characterize the sequence-dependent cross-bridging by a single fitting parameter, correlated to the number of bridges and their locations. At higher protein concentrations, we show that a polymer-like behavior is recovered, and mean-field models of the protein mechanical response are valid. 

\begin{figure}
\includegraphics[width=0.50\textwidth]{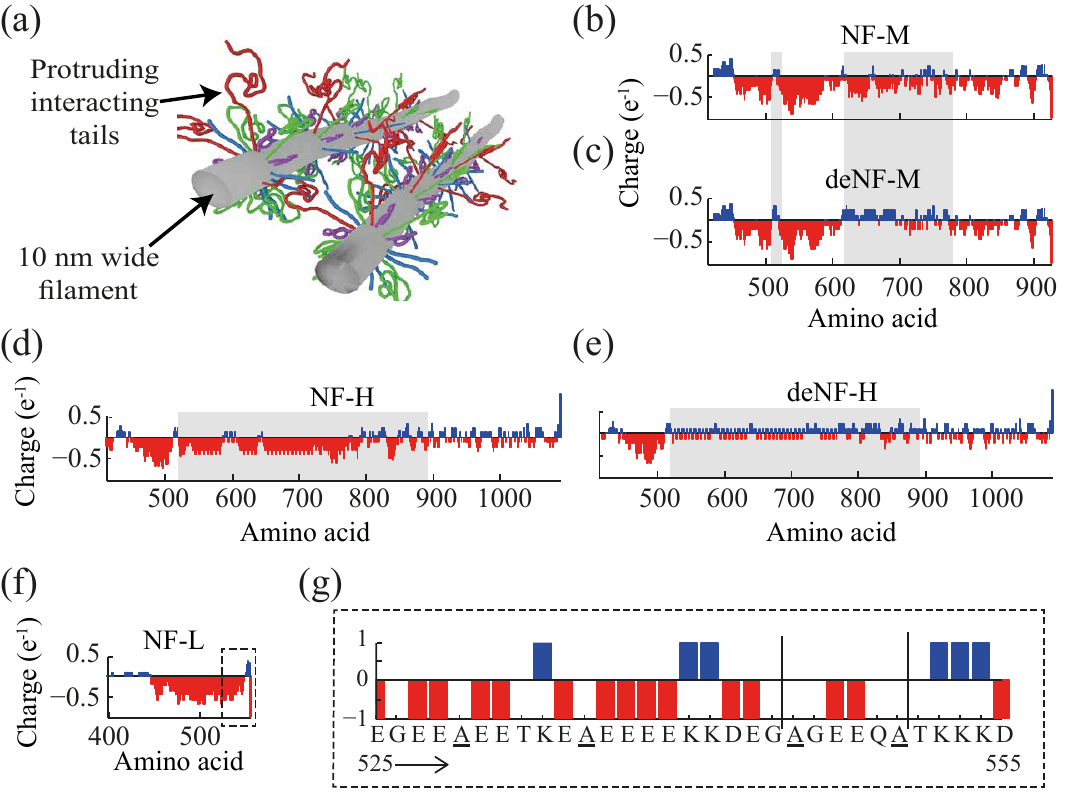}
\caption{(a) Neurofilament protein form bottlebrush filaments which interact via their protruding tails. Each filament is decorated with either one (NF-L), two (NF-L with NF-M or NF-L with NF-H) or three different types of tails. Tail charge distributions of native (b) NF-M and (d) NF-H,
are compared against the charge distributions of de-phosphorylated (c) NF-M, (e) NF-H and (f) NF-L. Charges are averaged over a 5-amino acid window at $p$H 6.8 (see Supplemental Material \cite{Note1}). (g) A close-up of NF-L
tip region shows the non-averaged amino acid charges, with two vertical
black line denoting the 5 and 11 amino acids truncated versions. Hydrophobic
residues are underlined.
 \label{fig:Intro}}
\end{figure}

The disordered proteins studied here, Neurofilament (NF) proteins, act as the stress-responsive skeleton of axons \cite{Azogui2015}. They self-assemble into a bottlebrush filament with 10 nm structured core decorated with the long disordered domains [Fig. \ref{fig:Intro}(a)]. At high concentrations, these  filaments condense into a hydrogel network whose properties are governed by interactions of the decorating disordered domains, known as tails \cite{Jones2008,Beck2010,Kornreich2015,Janmey2003,Kumar2002a,Beck2012}. We study filaments consisted of combinations of three different subunit proteins: NF-L, NF-M and NF-H. Notably, these proteins greatly differ in tail's length, net charge and charge distribution  [Fig. \ref{fig:Intro}(b-f)]. 


Below, we will show that at physiological conditions, the network's resistance to compression is primarily determined by steric interactions. We will argue that it is the lack of a fixed structure which allows the network to sustain considerable deformation, over 20-fold in volume, without yielding. In contrast, under little or no compression, we will experimentally demonstrate that short-ranged attractive electrostatic interactions determine the network expansion. The attractive motifs and polymeric repulsive forces offer a physical rationale for using disordered proteins in biological scaffolds. 

The NF protein purification and sequential assembly into hydrogel are detailed in ref. \cite{Kornreich2015} and the Supplemental Material \footnote{See Supplemental Material at \url{http://journals.aps.org/prl/}, which includes Refs. \cite{TheUniProtConsortium2014,Hancock2004,Harpaz1994,Kyte1982,Gibson2009,Mucke2004,DeGennes1980,Alexander1977a,Herrmann2004,Dobrynin2004,Onsager1949,Dobrynin2005a,Srinivasan2014a,Yasar2014}}. Following previous studies which predicated attractive interactions involving NF-L tail-end \cite{Kornreich2015}, we construct two new NF-L variants, NF-L5 and NF-L11. Their sequences are identical to NF-L, apart from the last 5 or 11 C-terminal truncated amino acids respectively [Fig. \ref{fig:Intro}(f,g)]. Further modification is induced by dephosphorylation of NF-M and NF-H, which significantly reduces the negative charge of their tails [Fig. \ref{fig:Intro}]. \nocite{TheUniProtConsortium2014,Hancock2004,Harpaz1994,Kyte1982,Gibson2009,Mucke2004,DeGennes1980,Alexander1977a,Herrmann2004,Dobrynin2004,Onsager1949,Dobrynin2005a,Srinivasan2014a,Yasar2014} 

We characterize the dominant tail interactions by measuring the structural properties NF hydrogels under osmolyte-induced osmotic pressure ($\Pi$). Equilibrated samples are measured by synchrotron small-angle X-ray scattering (SAXS) and cross-polarizing microscopy (CPM). SAXS provide direct structural measurement of the inter-filament distance ($D$) while CPM characterizes the macroscopic alignment of the filaments to nematic liquid crystals [Figs. S2 and S3 in the Supplemental Material \cite{Note1} as well as Refs. \cite{Jones2008,Beck2010,Kornreich2015,parsegian198629}].

\begin{figure}
\includegraphics[width=0.5\textwidth,draft=false]{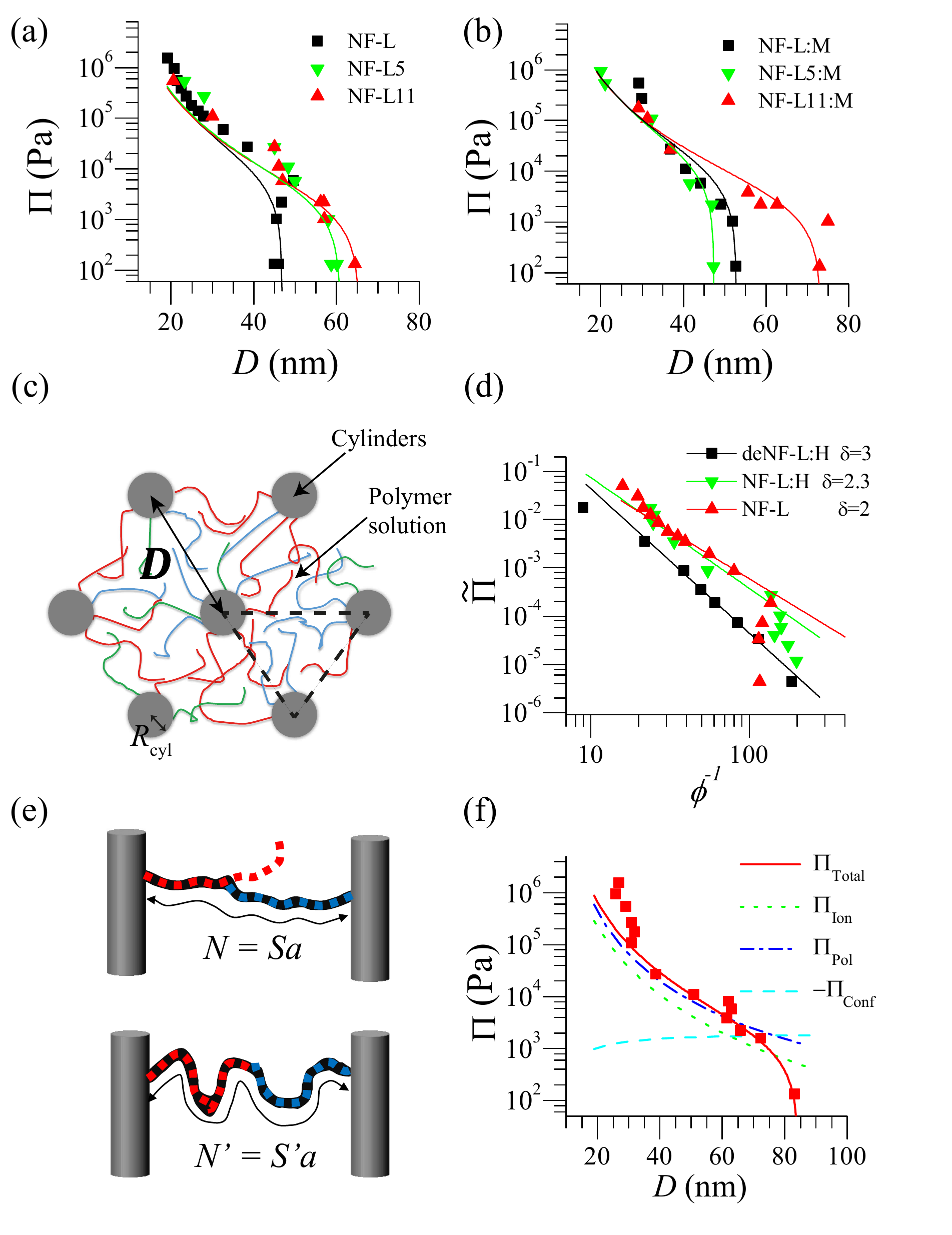}
\caption{(a) $\Pi$ vs. $D$ curve of NF-L and NF-L mutants at 150 mM show that truncation
of the positive short tip segment results with a $\sim$15 nm
expansion at lowest $\Pi$. (b) A similar
20 nm expansion is observed in the NF-L:M network as a result
of 11 amino acid tip truncation. (c) Sketch of the hexagonal model used
for the theoretical calculations, with the equilateral triangular unit cell of volume $V$ marked with dashed line. (d) Dimensionless osmotic pressure
$\tilde{\Pi}$ vs. $\phi^{-1}$ of different brushes exhibit a power law decay above $\phi\sim0.01$. (e) A schematic illustration of opposite filaments interconnected by an effective contour length $S=Na$, shown in thick solid black line. The location of tail-bridging segments sets $S$, and consequently, the attractive force prefactor $k$. (f) Modeled $\Pi$ terms are plotted against experimental NF-L:H data (\textcolor{red}{$\blacksquare$}). Fitted $\Pi$ are plotted as solid lines in (a,b).  \label{fig:P-D high salt figures}}
\end{figure}

The compression data, $\Pi$ vs. $D$, of homo- and hetero-polymer hydrogels reveals two distinct regimes [Fig. \ref{fig:P-D high salt figures}(a,b)].  Below $\Pi\sim10^4$ Pa the inter-filament spacing is divergent, as we also observed
for a much larger set of neurofilament networks [Fig. S5 in the Supplemental Material \cite{Note1}]. Importantly, replacing NF-L with its tip-truncated variants has a dramatic effect at this regime both on the homopolymer NF-L network and on the composite NF-L:M network. 

In the homopolymer (\emph{i.e.}, containing one tail type) NF-L network, the small tip truncation alters both the inter-filament
spacing and the macro-scale orientation [Figs. \ref{fig:P-D high salt figures}(a,b) and S2], as truncated NF-L variants produce more \emph{expanded} and isotropic networks.
A similar expansion due to NF-L truncation occurs in
the composite NF-L:M network. The truncation of 11 amino produced a
$\sim20$ nm expansion of the elongated NF-M tail. However, the hydrogel orientation was not affected, which remained nematic. These results validate our previous sequence-based studies, that predicted a specifically critical role of the NF-L tail end segment in setting the macroscopic properties of the hydrogel \cite{Beck2010,Kornreich2015}. 

In contrast, at $\Pi$ exceeding $\sim10^{4}$ Pa, networks share a similar compression
trend, irrespective of their length and charge fraction. To theoretically study their stress response under these compressions, we employ a mean-field approach, which is less sensitive to sequence variations. Since the networks are oriented under large deformations [Fig. S2], we treat the filaments as infinitely-long impenetrable cylinders of radius $R_{\mathrm{cyl}}=5$ nm. These are set in a hexagonal lattice [Fig. \ref{fig:P-D high salt figures}(c)], while the tails in-between the filaments act as a semi-dilute polymer solution \cite{Zhulina1995}. 

To validate the applicability of the approximation, we first compare the experimental data against
scaling laws of semi-dilute polymer solutions. We use an expanded $\Pi$ vs. $D$ dataset of NF-L, NF-M and NF-H which also includes data from refs. \cite{Beck2010,Kornreich2015,Eti2016}
at high salt concentrations (150 mM and above). Their dimensionless osmotic
pressure $\tilde{\Pi}=\Pi V_{a}/\left(k_{B}T\right)$ is plotted against
the tail inverse volume fraction $\phi^{-1}$ [Fig. \ref{fig:P-D high salt figures}(d)]. Here, $V_a=0.134$ nm$^3$ is the average amino acid volume, and $k_B T$ is the thermal energy taken at room temperature. The volume fraction is given by $\phi=NV_a/V$, where $V$ is the unit cell volume [Fig. \ref{fig:P-D high salt figures}(c,d) as well as Fig. S4 in the Supplemental Material \cite{Note1}] and $N$ is the number of tail amino acids in $V$. We find that at high osmotic pressure and high $\phi$ , the experimental data fits a power law decay with  $\tilde{\Pi}\propto \phi^{-\delta}$.  Best fits are obtained for $\delta\sim2-3$, in agreement
with known semi-dilute solution scaling laws \cite{RubinsteinM.2003,Dobrynin2005}. %

Using the lattice model, the osmotic pressure $\Pi$ is approximated by:
\begin{equation}
\Pi=\Pi_{\mathrm{ion}}+\Pi_{\mathrm{pol}}+\Pi_{\mathrm{conf}}.
\end{equation}
Here,  $\Pi_{\mathrm{pol}}$ is the interaction between protein segments;
$\Pi_{\mathrm{ion}}$ is the ionic osmotic pressure; and $\Pi_{\mathrm{conf}}$
will be fitted to our data to account for the non mean-field trend observed at low $\Pi$. We will show that $\Pi_{\mathrm{conf}}$ introduces an entropically driven force into the system, and that its magnitude is sequence specific. A detailed derivation of the three terms and their applicability is given in the Supplemental Material \cite{Note1}. 

$\Pi_{\mathrm{ion}}$ is calculated according to the Donnan rule \cite{Zhulina1995}:
\begin{equation} 
\Pi_{\mathrm{ion}}=k_{B}T\left(\sqrt{Q^{2}+4C_{s}^{2}}-2C_{s}\right),
\end{equation}
where $Q$ is the brush immobilized charge concentration and $C_{s}$ is the monovalent salt concentration in the reservoir. To calculate Q, we sum over the different tail types:
\begin{equation} 
Q=n_T \sum_{i=1}^{3}m_i\alpha_i c_{i,p}. 
\end{equation}
For a given tail type $i$, $c_{i,p}$ and $\alpha_{i}$ are the amino acid concentration and charged amino acid fraction, respectively; the total number of tails in a unit cell is $n_T$=16; the tail type molar fraction $m_i$ holds $\sum_{i=1}^{3}m_i=1$.

For $\Pi_{\mathrm{pol}}$, we treat polymer segments as a non-ideal
gas with virial corrections. Good
 agreement with experiments is achieved by using a thin-rods gas model:
\begin{equation}
\Pi_{\mathrm{pol}}=k_{B}T\left(B_{2}\phi^{2}+B_{3}\phi^{3}\right).
\end{equation}
The entropic term, which depicts the translational entropy of free polymers and is linear in $\phi$, is omitted since the tails are connected to the backbone. The quadratic term, $B_2$, accounts for the pair-wise excluded volume interaction between polymer rod segments. The average amino acid length along the polypeptide tail is $a=0.35$ nm. Each rod is one Kuhn-length ($l_k$) long, and therefore each tail $i$ has $N_{k,i}=N_{i}/\left(l_{k}/a\right)$ statistical Kuhn segments. We assume segments to be of the order of one Kuhn length $k\sim2l_{p}$ \cite{Grosberg1994}, where the persistence length $l_{p}$ of unstructured proteins is in the range $0.4-0.8$ nm \cite{Hofmann2012,Cheng2010}. The third virial term, $B_3$, is negligible within the concentrations used here \cite{Straley1973}. 

Both $\Pi_{\mathrm{pol}}$ and $\Pi_{\mathrm{ion}}$ are set by the experimental conditions ($e.g.$, buffer salinity, $p$H) and NF protein stoichiometry without any fitting parameters. However, the attraction between the tails is governed by local interactions that bridge adjacent filaments. We suggest a Flory entropic term \cite{RubinsteinM.2003}, to account for the bridging attraction interaction: 
\begin{equation}
\label{Eq_conf}
\Pi_{\mathrm{conf}}=-k\frac{\partial}{\partial V}\left[k_{B}T n_T\sum\limits_{i=1}^{3}m_i\frac{3\left(D/2-R_{\mathrm{cyl}}\right)^{2}}{2N_{k,i}l_{k}^{2}}\right].
\end{equation}
The one fitting parameter, $k$, semi-quantitativly captures the dependence of the attractive polymer spring on the bridging locations and multiplicity [Fig. \ref{fig:P-D high salt figures}(e)]. The assumption that the entropic term is attractive is supported post-priori by the good agreement of the fits to the experimental data. Furthermore, it conceptually replaces the attractive term which appears in the Alexander-de Gennes model for neutral planar brushes \cite{DeGennes1987}. 

The attractive force can be rationalized by the effective contour length ($S$) connecting adjacent filaments via an entropic polymer spring [solid line in Fig. \ref{fig:P-D high salt figures}(e)]. A pair of tails may form different connecting contour lengths, depending on their bridging locations. For example, if each tail-end forms an inter-filament cross-bridge with an identical tail, without inter-penetration, then $S$ is maximal and $k$ approaches unity. When the brush is not uniform and fewer tails stretch out to $D/2$, fewer connections exist and $k$ would decrease. Alternatively, if tails cross-bridge at segments closer to filament backbone (along the polypeptide chain), the effective contour would decrease and $k$ would increase. 

\begin{table*}[ht]
\caption{\label{tab:table5} Monomer charge fraction $\alpha=\sum_{i=1}^{3}m_i \alpha_i$ at $p$H 6.8, fitted spring ``constant'' $k$ parameter, and the hydrogel orientation (Nematic or Isotropic) of NF networks at low $\Pi$ and 150 mM monovalent salt. NF-L tip truncation is followed by $k$ decrease due to the tip's key role in bridging interactions. Similarly, removal of the negatively charged phosphates from NF-M debilitates the cross-linking. See Supplemental Material Table S2 for protein molar fractions, Fig. S2 for CPM images as well as Fig. S7 on the conservation of the positively charged NF-L tip in different species \cite{Note1}.}
\begin{ruledtabular}
\begin{tabular}{ c c c c c c c c c c c c}
	Composition & L               & L5              & L11               & L:M             & L5:M & L11:M & de(L:M) & L:H  & de(L:H) & L:M:H & de(L:M:H) \\ \hline
	$\alpha$    & 0.26            & 0.28            & 0.28              & 0.24            & 0.25 & 0.25  & 0.20    & 0.24 & 0.20    & 0.23  & 0.19      \\
	$k$         & 0.55            & 0.17            & 0.11              & 0.92            & 1.46 & 0.20  & 0.23    & 0.10 & 0.19    & 0.20  & 0.33      \\ 
	Orientation & N               & N               & I                 & N               & N    & N     & N       & N    & I       & N     & N
\end{tabular}
\end{ruledtabular}
\end{table*}

At high salt concentrations, our approach agrees well with the experimental data over a large range of osmotic pressures at
multiple subunit stoichiometries and phosphorlaytion states [Figs. \ref{fig:P-D high salt figures}(a,b)
and S5]. \emph{The single fitting parameter} ($k$) affects the attractive
term only ($\Pi_{\mathrm{conf}}$), which is significant at the low pressure regime [Fig. \ref{fig:P-D high salt figures}(f)]. 

At $\Pi>10^4$ Pa, the two repulsive terms are dominant. In most cases, $\Pi_{\mathrm{pol}}>\Pi_{\mathrm{ion}}$ but not by more than a magnitude [Fig. \ref{fig:P-D high salt figures}f]. Therefore, the electrostatic repulsion plays only a secondary role in setting the hydrogel compression response at physiological conditions (150 mM). Under these large deformations, the tails are successfully treated as a non-grafted semi-dilute polymer solution [see also expanded dataset in Fig. S5]. 

\begin{figure}
\includegraphics[width=0.5\textwidth]{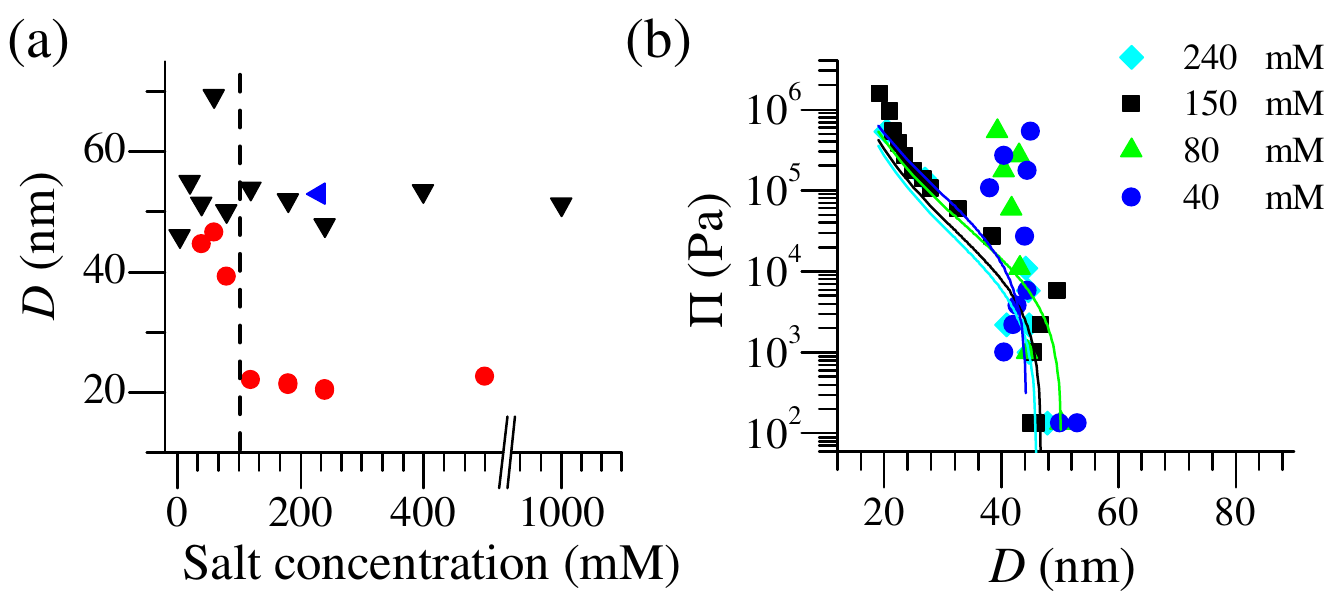}
\caption{(a) NF-L inter-filament spacing vs. monovalent salt concentration of NF-L at: 0 \% PEG ($\blacktriangledown$), 20 \% PEG (\textcolor{red}{$\bullet$}), and EDTA assembly buffer with 0 \% PEG (\textcolor{blue}{$\blacktriangleleft$}). Increasing concentrations beyond 100mM do not significantly modify the stress response, in agreement with previous simulations \cite{Jayanthi2013}. The inter-filament distance measured in NF-L networks formed with EDTA buffer show that the bridging mechanism does not require divalent salts. (b)   $\Pi$ vs. $D$ results of NF-L at a wider range of monovalent concentration shows significant deviations from the model at lower salt concentrations. \label{fig:LowerSalt}}
\end{figure}

At lower $\Pi$, the attractive term $\Pi_{\rm{conf}}$ is non-negligible. Since this term is sequence-dependent, even minute charge modifications can be followed by dramatic macroscopic effects, as seen in the NF-L truncations [fig. \ref{fig:P-D high salt figures}(a,b)]. In this case, interactions of the NF-L tip with tail segments closer to the filament backbone may have forced the native NF-L into a collapsed loop-like conformation \cite{Jeong2014a}.

Another evidence for the role of charge distribution in setting the attractive force is provided
by the effect of 50\% charge removal off the NF-M brush by dephospohrylation. The charge reduction results with a significant $20$ nm network \emph{expansion}, opposite to the naive electrostatic expected trend. Again, it is captured by the parameter $k$ which is related to the sequence organization and not to the tails' charged fraction (Table \ref{tab:table5}). 

At lower salt concentrations the model fails to predict the increase
in osmotic pressure, Figs. \ref{fig:LowerSalt}(b,c) and S6. The power law decay of NF-L at 80 mM is fitted with $\delta=8$, which implies that the semi-dilute
approach is not valid here. The unusually high exponent may be explained
if one assumes that the tails are well-extended and that the excess
pressure builds against the stretched tails. Such scenario is
reminiscent of  covalent bonds and agrees with the known non-linear stress response of IFs. There, the change of the differential modulus with stress follows a $3/2$ exponent, typical of covalent cross-linked gels \cite{Pawelzyk2014}. Since no covalent binding sites are found on NF tails, this indicates a significant cross-bridging energy \cite{Beck2010,Kornreich2015}. 

In contrast, the larger spacing measured for NF-L5 suggests that NF-L
is not overly stretched at $D=50$ nm. Nonetheless, this counter-argument
does not necessarily overrule the ``stretching hypothesis'' since
NF-L5 and NF-L11 filaments are also less aligned [Fig. S2]. Moreover, best agreement with low-salt data is achieved for the  NF-L:M hydrogel at 70 mM, which is well-oriented even at low $\Pi$. Such relation between orientation and the hydrogel properties require further theoretical investigation \cite{Deek2013,Deek2016,Leermakers2008}.


In summary, NFs form a hydrogel whose mechanical and structural properties are governed by a semi-dilute solution of long tails. We successfully model the hydrogel mechanics at near physiological conditions as a non-ideal gas of steric cylinders and small ions. Accordingly, the system acts
as a pneumatic shock absorber in equilibrium, where the ionic and steric gas-like repulsive forces build pressure against external forces.
Such mechanical behavior suggests a novel rationale for the use of disordered proteins as essential building blocks in the cellular supporting framework. 

The main advantage of pneumatic over rigid devices, is that the former is less subjected to
damage under large deformations. The osmotic response of the disordered proteins in the biological hydrated environment, resembles that of a pressurized gas. This is especially important in the axon, where large organelle transport and axonal flexibility require both a soft and elastic scaffold \cite{Safinya2015}. The
employment of transient (attachable-detachable) cross-links, as demonstrated here, enables the deformation and reformation of the network \cite{Monsma2014,Lieleg2008}. The binding is facilitated by the large conformational space explored by the disordered tails.

A comparison with other cytoskeletal elements points out that disordered protein decoration of filaments in cellular scaffolds is widespread \cite{Srinivasan2012}. 
NF proteins are part of a large group of over 70 cell-specific \emph{disordered} proteins, which provide different cells with their unique mechanical scaffolds \cite{Kornreich2015a}. Another major cytoskeletal element, microtubules, is decorated by disordered proteins which was also shown to mediate microtubules inter-filament interactions \cite{Chung2015,Chung2016,Mukhopadhyay2001}. As similar environmental conditions prevail in all these systems, we expect that the physical principles laid here would also apply for this greater family.

\begin{acknowledgments}
{\bf Acknowledgments.~~~} {We are grateful to Dr. Geraisy Wassim of Beit Shean abattoirs Tnuva for kindly providing us with the spinal cords. We thank Ekaterina Zhulina, Philip Pincus, Yacov Kantor, Tomer Markovich, Michael Rubinstein and Haim Diamant for useful discussions and suggestions. We thank the following beamlines for SAXS measurements: I911-SAXS at MAX IV Laboratory, Lund, Sweden; SWING at SOLEIL synchrotron, Paris, France; and I-22 beamline at
Diamond, England. This work was supported by Israeli Scientific  Foundation (571/11, 550/15), the Tel Aviv University Center for Nanoscience and Nanotechnology and the Scakler Institute for Biophysics at Tel Aviv University. Travel grants to synchrotron facilities were provided by BioStruct-X. }
\end{acknowledgments}
\bibliographystyle{apsrev4-1}
\bibliography{library}

\pagebreak

\end{document}